# Thermal annealing effects on Graphene/n-Si Schottky junction Solar cell: Removal of PMMA residues

Yuzuki Ono, Hojun Im[*]

*Graduate School of Science and Technology, Hirosaki University, Hirosaki 036-8561, Japan*

**ABSTRACT**

Thermal annealing is one of most effective way to improve the efficiency of graphene/n-Si Schottky junction solar cell. Here, its underlying mechanism has been investigated by comparative studies in terms of the removal of polymethyl methacrylate (PMMA) residues, using the *J-V* characteristics, the transient photocurrent and photovoltage measurements. Experimental results have revealed that there are trap states which are originated from the PMMA residues and cause the large photocurrent leakage as the intensity of the incident light increases. It is also found that the PMMA residues accelerate deterioration and rapidly invalidate hole doping effects. Such undesirable PMMA residues were effectively removed by the thermal annealing treatments, serving to reduce the photocurrent leakage and to increase the stability.

[*]Corresponding author. Email: hojun@hirosaki-u.ac.jp (H. J. Im)

## 1. Introduction

Graphene/n-Si Schottky junction has attracted extensive attention for applications of next generation optoelectrical devices such as solar cell, photodetector, photosensor, utilizing their unique electrical and optical properties [1–5]. Among them, graphene/n-Si Schottky junction solar cell (GSSC) has shown remarkable developments [6–12]. Since the first report in 2010, its power conversion efficiency (*PCE*) has increased rapidly from ~1.5 to ~15% by successful efforts such as hole doping of graphene, passivation of silicon surface, anti-reflection coating, etc [13–22]. However, there have still several issues to overcome toward practical applications. In the fabrication process of GSSC, polymethyl methacrylate (PMMA) has been widely used as a support layer during transfer of graphene onto a target such as n-Si substrate [23–25]. After the transfer process, the PMMA layer has been removed usually by acetone. However, it has been hard to completely remove PMMA, hence its residues have caused the degradation of *PCE* [26–30]. Thermal annealing treatment has been considered as one of the most useful methods to remove the PMMA residues and performed usually in the forming-gas and/or vacuum [24,30], enhancing the performance of GSSC [28,31]. To fully utilize such thermal annealing effects, it is inevitable to understand their underlying mechanism

Here, we have studied the thermal annealing effects on GSSC by the current density vs. the bias voltage (*J-V*) characteristics and the transient photocurrent and photovoltage (TPC/TPV) measurements, focusing on the removal of the PMMA residues. The obtained experimental results have revealed that there are the trap states originated from the PMMA residues, which cause the retardation of the photocurrent transport and the photocurrent leakage. In addition, the PMMA residues rapidly invalidate the hole doping effects and accelerate the degradation of its performance. In comparison between GSSCs with and without annealing, we observed that the thermal annealing treatments effectively remove such undesirable PMMA residues, reducing the photocurrent leakage, increasing the stability, and persisting the hole dop effects.

## 2. Experimental details

### 2.1. Graphene growth

Graphene was grown on the catalytic substrate of 30 μm-thick Cu foil by the low-pressure chemical vapor deposition (CVD). The Cu foil was annealed at 980 °C for 1 h in forming gas ($H_2$:Ar = 1:9) at ~10 mbar. Subsequently, the mixed gases of $CH_4$ (5 sccm) and $H_2$ (30 sccm), which are play roles as the precursor and reduction gas, respectively, were introduced into CVD quartz chamber for 30 min at 980 °C and ~10 mbar, followed by cooling down the chamber by full-open of the CVD furnace. Graphene was transferred to a target substrate by using the conventional PMMA assisted method. PMMA was spin-coated on the graphene/Cu-foil layer at 3000 rpm for 60 s, and was cured at 80 °C for 10 min. After the removal of graphene grown on the backside of Cu foil by $HNO_3$ (10 wt%) etching for 1 h, Cu-foil was completely etched off by $FeCl_3$ solution of 2.5 wt% for 3 h.

### 2.2. GSSC fabrication

n-type (100) Si wafer of 200 μm thickness and 1 - 10 Ω with thermal $SiO_2$ layers (~ 500 nm) in both front and back sides was used as a device substrate. The thermal oxide layers were removed using buffered oxide etchant for 20 min, followed by cleaning process with acetone, isopropyl alcohol, deionized water, and dry nitrogen gas. Active area was defined as a window of 3×3 $mm^2$ by photolithography. PMMA/graphene was transferred onto the patterned n-Si substrate by the conventional fishing method. In this work, we have prepared two types of GSSCs. One type is a solar cell where the PMMA layer was removed by the acetone for 90 min. In the other type device, the PMMA residue was additionally removed by thermal annealing treatment in forming gas ($H_2$:Ar = 1:9) at ~420 °C at near 1 atm. Finally, the solar cells have been completed by forming the back and front contacts using GaIn eutectic and silver paste, respectively.

### 2.3. Characterization

The *J-V* characteristics were measured using a source meter (Keithley2400, Tektronix) in dark and under illumination. We used a xenon lamp with AM1.5G filter as the light source of 1 sun (100 mW/cm$^2$). In light-intensity dependent measurements, a halogen lamp (MegaLight100, Schott) was employed. The light intensity was adjusted by a NIST traceable calibrated Si-photodetector (FDS1010, Thorlabs). The TPC and TPV responses were obtained using a laser diode of 639 nm (HL6358MG, Thorlabs) and an oscilloscope (MSO5354, Rigol). External quantum efficiency (EQE) was measured by home-built equipment with a monochromator (CS260, Newport) and lock-in amplifier (LI5640, NF) using the halogen lamp.

## 3. Results and discussion

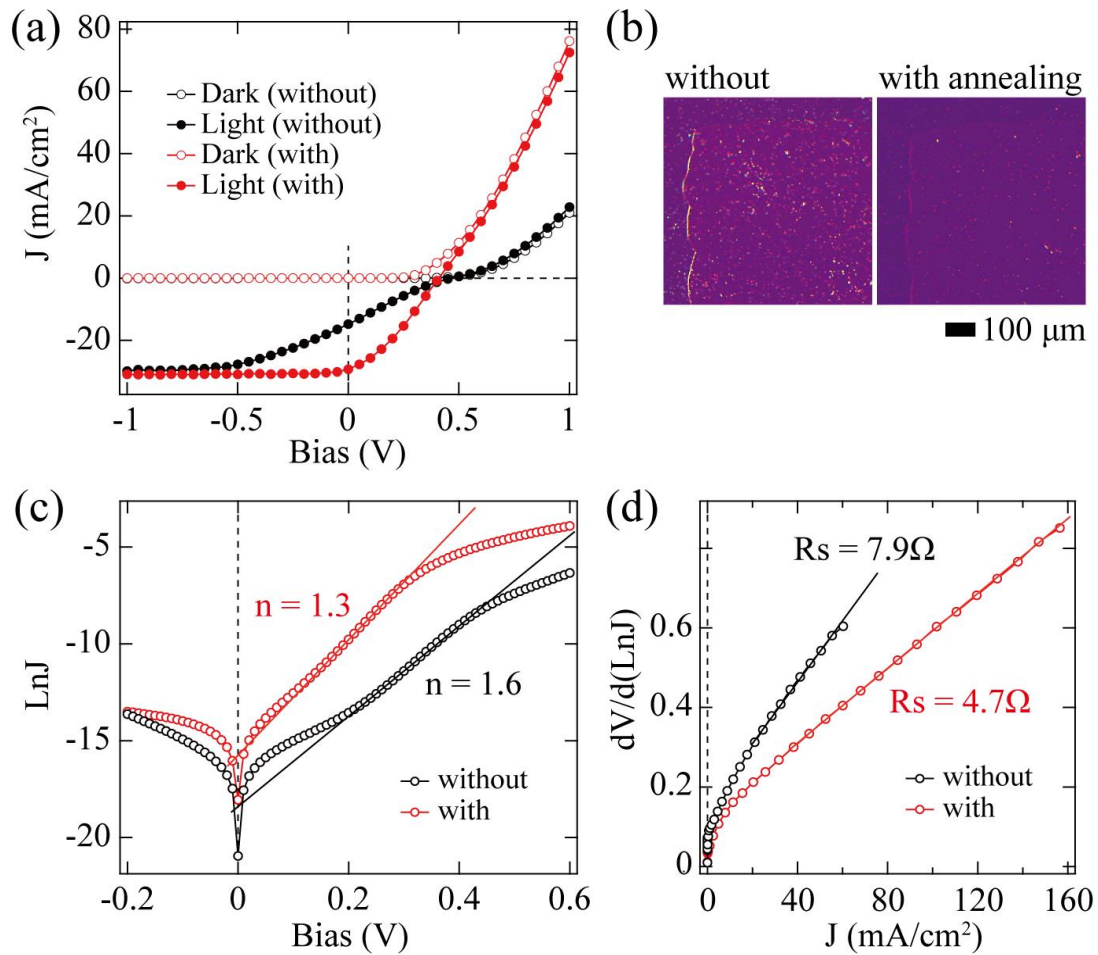

**Fig. 1.** (a) *J-V* characteristics and (b) photographs of the w-a and wo-a GSSCs. Plots of (c) $\ln J$ vs. $V$ and (d) $dV/d\ln J$ vs. $J$, corresponding to the *J-V* characteristics. The solid lines are guide to the eye.

The top views of the active area of GSSCs without and with the thermal annealing treatment are shown in the left and right panels of Fig. 1(b), respectively. In the without-annealing (wo-a) GSSC, we can observe small white spots with size of several μm, which have been considered to mainly come from PMMA residues after removal by acetone [27,30,32]. On the other hand, in the with annealing (w-a) GSSC such PMMA residues were effectively eliminated by thermal annealing treatment, even though it is not perfect. The dark *J-V* curves of both GSSCs show the excellent rectifying properties, indicating that the Schottky junction well formed between graphene and n-Si substrate (Fig. 1(a)) [3,4]. By the analysis based on the thermionic-emission model [3,33,34], the ideality factor ($n$) and the Schottky barrier height (*SBH*) from the $\ln J$ vs. $V$ plots were estimated to be 1.3 (1.6) and 0.82 (0.89) eV for the w-a (wo-a) GSSCs, respectively (Fig. 1(c)). The series resistances ($R_S$) were obtained from the slope of $dV/d\ln J$ vs. $J$ as shown in Fig. 1(d) [21,35–37], being 4.7 (7.9) Ω for the w-a (wo-a) GSSCs, respectively. These indicate that the Schottky junction properties were improved by the thermal annealing treatments. Under the illumination of 1 sun, it is observed that the thermal annealing treatments also change the photovoltaic parameters, the short-circuit current ($J_{SC}$), open-circuit voltage ($V_{OC}$), fill factor (*FF*), and *PCE*, from 14.7 mA/cm$^2$, 0.47 V, 22 %, and 1.5 %, to 29.2 mA/cm$^2$, 0.41 V, 33 %, and 3.9 %, respectively. In addition, the s-shape feature around $V_{OC}$, observed in wo-a GSSC, disappeared in w-a GSSC. These also mean that the thermal annealing treatments play an effective role to enhance the performance of GSSCs in agreement with the previous papers [28]. The extracted photovoltaic parameters were summarized in table S1.

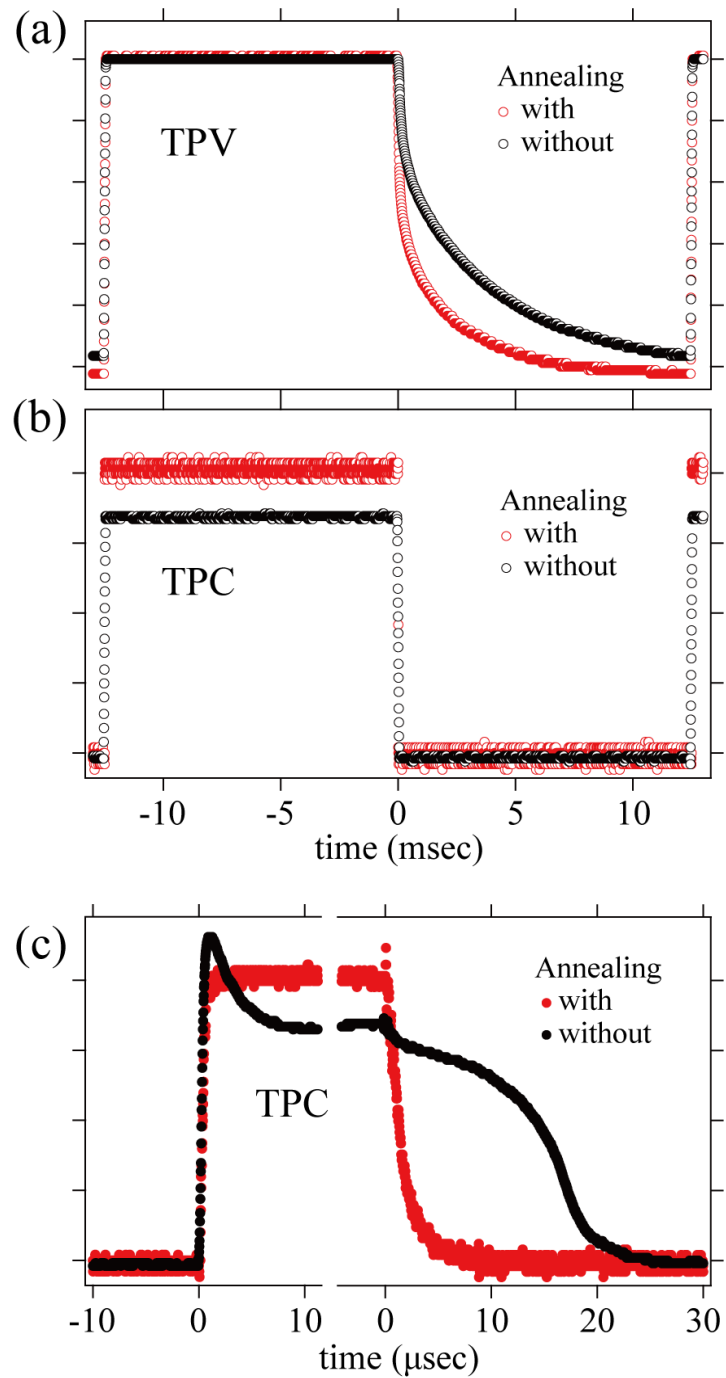

**Fig. 2.** (a) TPV and (b) TPC responses of the w-a and wo-a GSSCs. (c) Enlarged TPCs near the rising and falling edges. For the sake of comparison, the TPV responses are normalized to each its own maximum intensity, while the TPC responses are raw data.

To understand the dynamics of the thermal annealing effects, we have performed the TPV and TPC measurements using a square pulse of 639 nm. In Fig. 2(a), it recognized that the TPV of the wo-a GSSC more slowly decays than that of the w-a GSSC. In general, the relaxation time of TPV has been interpreted as the recombination time of exciton (photoexcited electron-hole pair); hence the longer relaxation time has a benefit of the efficiency of solar cell. However, the obtained TPV results show an opposite tendency to the usual interpretation and require a consideration of additional factors. For the detail analysis, we fitted the TPV responses by the double exponential decay equation ($Ae^{-i\omega\tau_1} + Be^{-i\omega\tau_2}$). The two relaxation times were estimated to be $\tau_1 = 1.1 \times 10^{-4}$ ($1.8 \times 10^{-4}$) s and $\tau_2 = 2.4 \times 10^{-3}$ ($4.2 \times 10^{-3}$) s for the w-a (wo-a) GSSCs, respectively. These values show good agreement with the previous reports, where $\tau_1$ and $\tau_2$ have been considered to come from the interface of Schottky junction and the Si bulk, respectively [14]. Although it is difficult to accurately evaluate the $\tau_1$ values due to too short time scale compared to the fitting region, we can recognize that there are not significant changes after the thermal annealing treatments [14]. On the other hand, the $\tau_2$ value of wo-a GSSC is apparently longer than that of w-a GSSC as shown in Fig. 2(a). Usually, thermal annealing at ~ 420 °C for 3 h is not sufficient to change the Si bulk properties. $\tau_2$ can thus be interpreted as the relaxation time which includes the voltage drop of the photogenerated carries in the PMMA residue as well as the Si bulk. Therefore, it should be noted that the increase of $\tau_2$ does not simply mean the enhancement of the GSSC performance.

Fig. 2(b) and 2(c) show the TPC responses of the w-a and wo-a GSSCs in the same time scale to TPV and in the enlarged time scale, respectively. The relaxation time of TPC has been regarded as a charge extraction time. The TPC of w-a GSSC shows a simple exponential decay in a few μs. This faster decay of TPC, compared to that of TPV, indicates that photogenerated current is effectively collected to the electrodes. On the other hand, in the TPC of wo-a GSSC, it is observed that there is the immediate decay of photocurrent by ~28% at the rising edge, causing the parasite current leakage which is possibly responsible for the low *FF* and *J$_{SC}$* in the

*J-V* characteristics (Fig. 1(a)). Furthermore, the shoulder is observed at the falling edge after the light turns off. This indicates that there are photocurrents retarded by a few 10 µs, which is a larger scale by one order of magnitude compared to that of the w-a GSSC and results in the increase of the recombination chance. These behaviors of TPC can thus be attributed to the photogenerated carriers trapped in the PMMA residues.

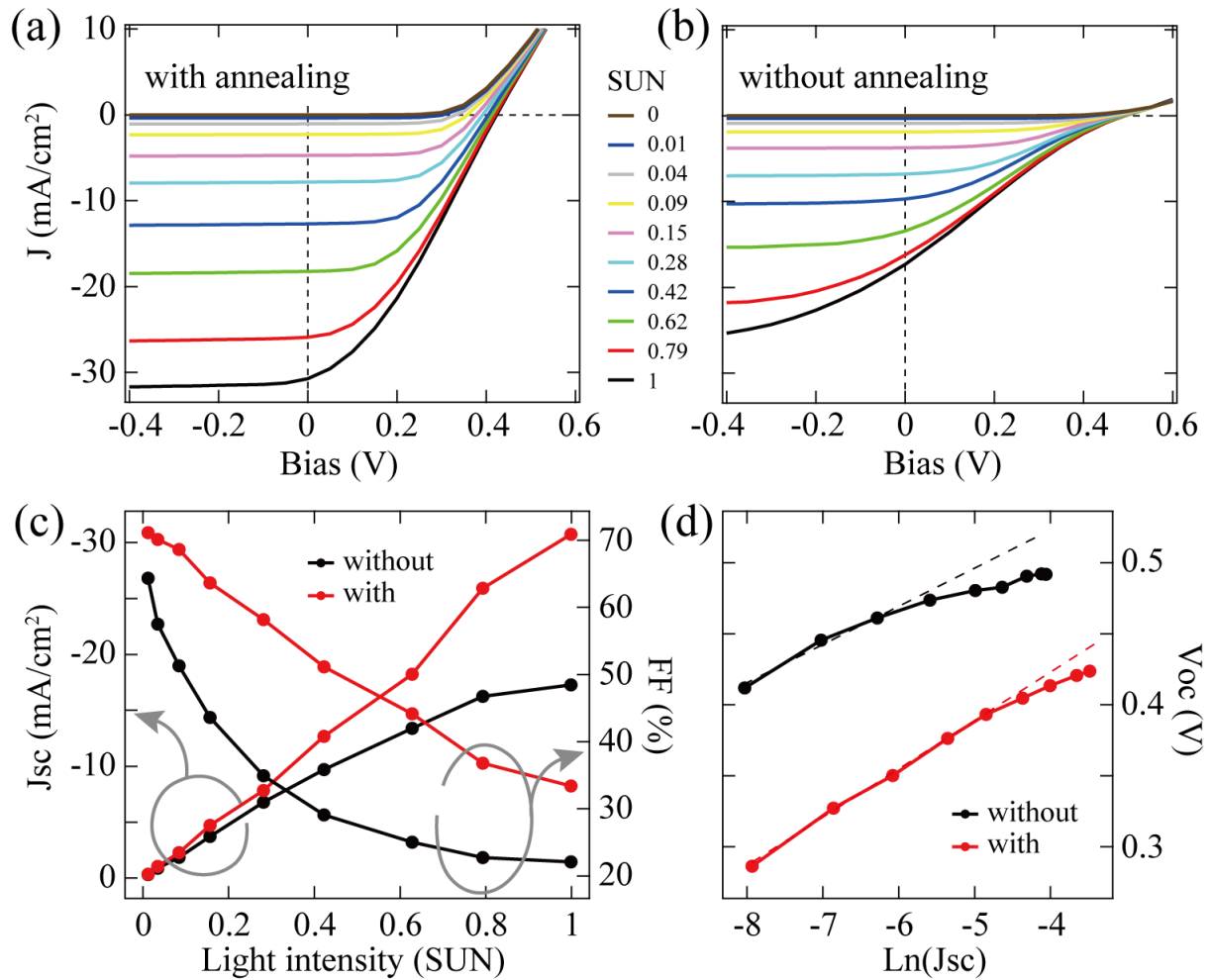

**Fig. 3.** Light intensity dependency of the *J-V* characteristics of (a) the w-a and (b) wo-a GSSCs. (c) Plots of $J_{SC}$ and $FF$ as a function of the light intensity. (d) Plot of $\ln J_{SC}$ vs. $V_{OC}$. The dashed lines are guide to the eye.

Figure S1 shows the EQE spectra of the w-a and wo-a GSSCs in the range of 350 – 1100 nm in agreement with the previous reports [14,15]. Despite the large difference of $J_{SC}$ values in *J*-

*V* measurements, both GSSCs show the similar spectral feature. When we consider that the EQE spectra are measured using the very weak light intensity compared to the intensity of the 1 sun, it can be easily inferred that there is a light-intensity dependency in solar cell performance.

To elucidate it, we have investigated the *J-V* characteristics of the w-a and wo-a GSSCs as a function of the light intensity (Fig. 3(a) and 3(b)). A halogen lamp, which can continuously adjust the light intensity, was used. The light intensity of the halogen lamp, where $J_{SC}$ has the same value in the 1 sun, was regarded as the 1 sun-equivalent condition. The variation of $J_{SC}$ and *FF* were plotted in Fig. 3(c). The $J_{SC}$ values of the w-a GSSC are linearly proportional to the light intensity. On the other hand, the $J_{SC}$ values of the wo-a GSSC monotonically increases with the light intensity in the weak light region but start to saturate as the light intensity closes to the 1 sun-equivalent condition. Similar behaviors are also seen in the variation of FF values, but inversely proportional to the light intensity. In Fig. 3(d), the $\ln J_{SC}$ vs. $V_{OC}$ relation of the w-a GSSC shows better linearity than that of the wo-a GSSC, indicating that the Schottky junction of the w-a GSSC is more ideal than that of the wo-a GSSC [33]. These results point out that, as the light intensity closes to the 1 sun, the photocurrent leakages occur and are larger in the wo-a GSSC than in the w-a GSSC. This also explains why the EQE spectra are similar in the w-a and wo-a GSSCs (Fig. S1). As a result, it is reasonable that the PMMA residues trap photogenerated carriers and deteriorate the performance of GSSC via the photocurrent leakage.

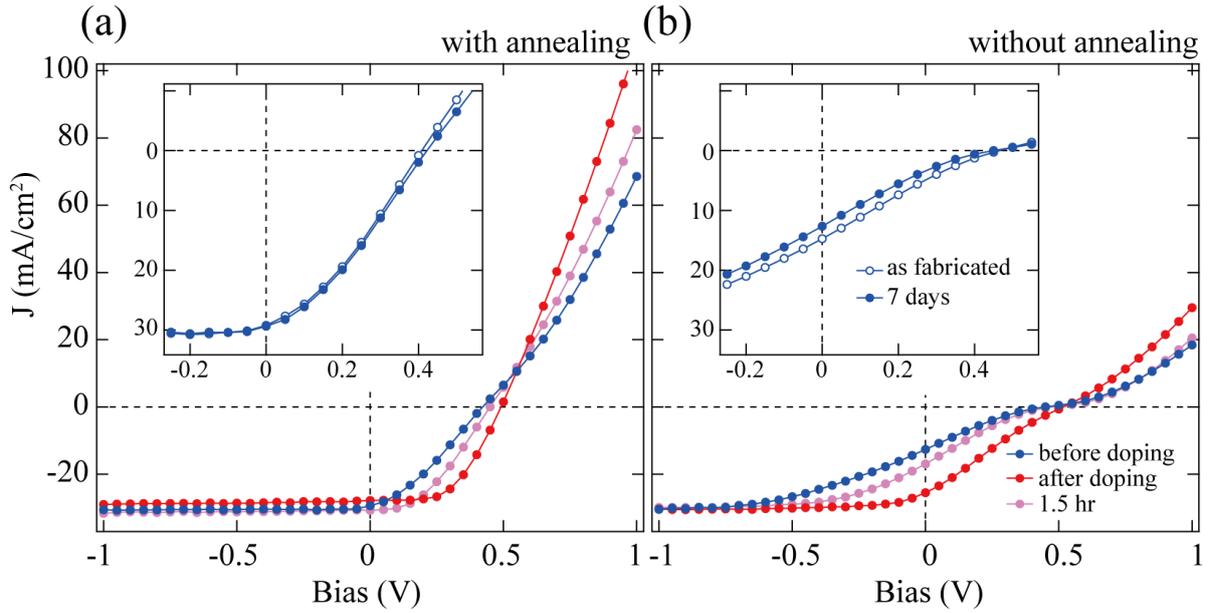

**Fig. 4.** The *J-V* characteristics of the w-a and wo-a GSSCs, hole doped by the HNO$_3$ evaporation, as a function of time. The insets show the *J-V* curves obtained soon and 7 days after the fabrication of devices before the hole doping.

Finally, we investigated the thermal annealing effects on the hole doping of graphene and its ageing in the w-a and wo-a GSSCs. The *J-V* characteristics of pristine and hole-doped GSSCs have been measured as a function of time (Fig. 4(a) and (b)). Degradation was not observed in the w-a GSSC for 7 days since the fabrication (the inset of Fig. 4(a)), while there was degradation of *PCE* from 1.5 to 1.2 % in the wo-a GSSC (the inset of Fig. 4(b)). After that, both GSSCs were hole doped by the evaporation of HNO$_3$ (69 wt%) for 1 min. It is well known that the hole dop of graphene lowers the Fermi level, improving the device performance [14,21]. The *PCE* values of the w-a (wo-a) GSSCs were enhanced from 4.0 (1.2) to 7.1 (2.5) %, respectively, soon after the hole doping. The related photovoltaic parameters were summarized in table S2 - S5. In addition, we recognize that the degradation of the wo-a GSSC is faster than that of the w-a GSSC; *PCE*s are reduced by ~21 (~40) % in 1.5 h after the hole doping for w-a (wo-a) GSSCs, respectively, indicating that the PMMA residues accelerate the deterioration of GSSC. In other words, the thermal annealing treatments have effects to enhance the stability as well as *PCE* by removing the PMMA residues. For reference, the light intensity dependencies of the *J-V* characteristics 1 h after the hole doping were also displayed in Fig. S2; the variation of $J_{SC}$ and *FF*, and ln$J_{SC}$ vs. $V_{OC}$ show the similar behaviors with those of the pristine GSSCs (Fig. 3).

## 4. Conclusion

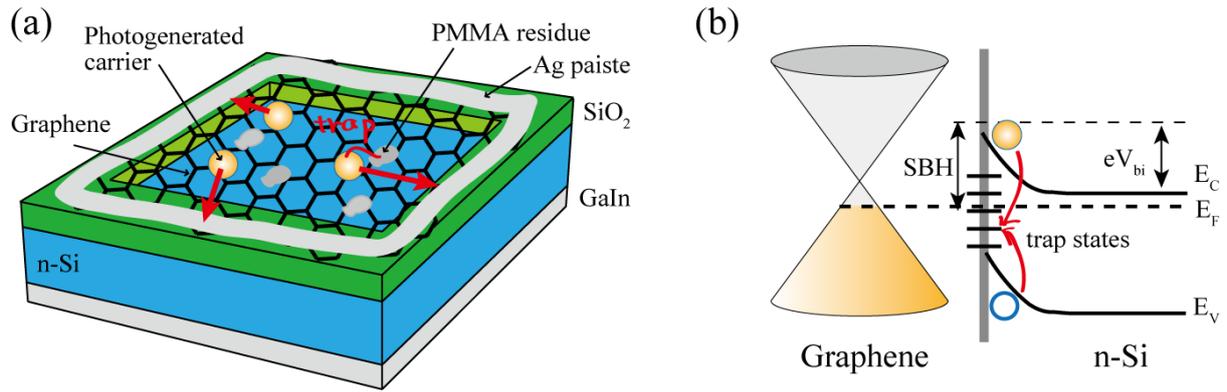

**Fig. 5.** (a) Schematic illustration and (b) energy diagram of GSSC with the PMMA residues. *SBH* and $V_{bi}$ represent Schottky barrier height and the built-in potential, respectively.

In summary, we have investigated the thermal annealing effects on GSSC in terms of the removal of the PMMA residues and elucidated its underlying mechanism by the comparative studies. In TPC measurements, there were the trap states originated from the PMMA residues, which cause the retardation of the photocurrent transport. In TPV measurements, the presence of the PMMA residues could be observed as the increase of the decay time. The light-intensity dependent *J-V* characteristics revealed that the PMMA residues give rise to the photocurrent leakage, which is a crucial factor to reduce the GSSC efficiency. It is also found that the PMMA residues rapidly invalidate the hole doping effects and accelerate the deterioration of GSSC. These results were depicted as schematic illustration and energy diagram in Fig. 5. From the comparison of the w-a and wo-a GSSCs, such undesirable PMMA residues were effectively removed by the thermal annealing treatments, serving to reduce the photocurrent leakage, to increase the stability, and to persist the hole dop effects. We believe that the revealed underlying mechanism of the thermal annealing effects, mainly related to the removal of the PMMA residues, is useful to make a strategy to design and improve GSSC. This also emphasizes an importance of the developments of several methods to effectively remove the PMMA residues.

SUPPLEMENTARY MATERIALS

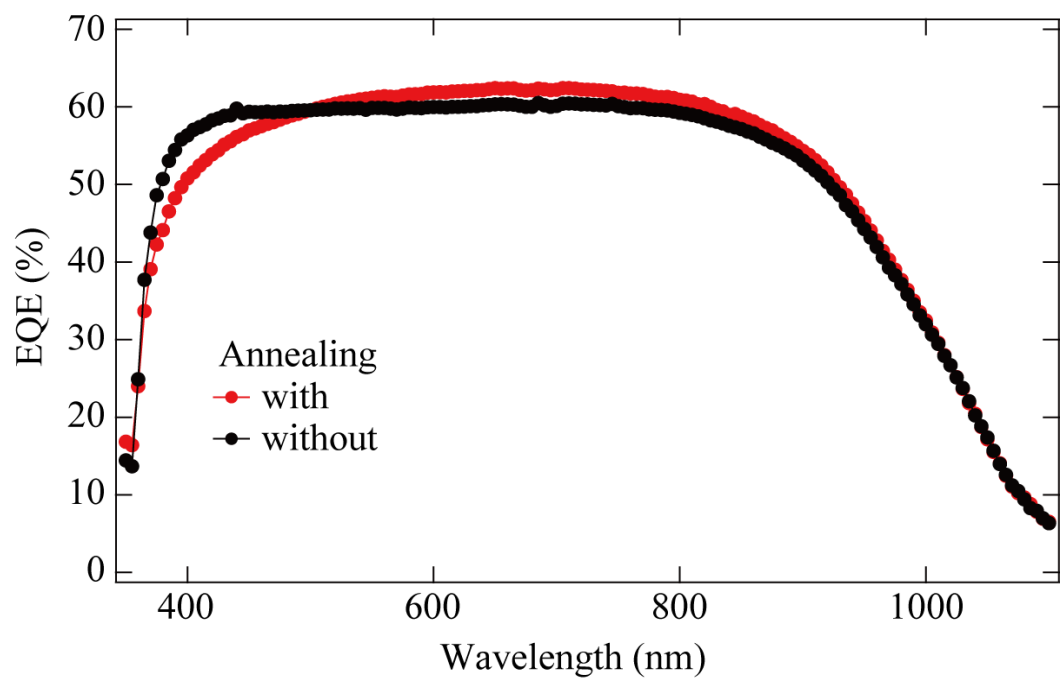

**Fig. S1.** EQE spectra of the w-a and wo-a GSSCs.

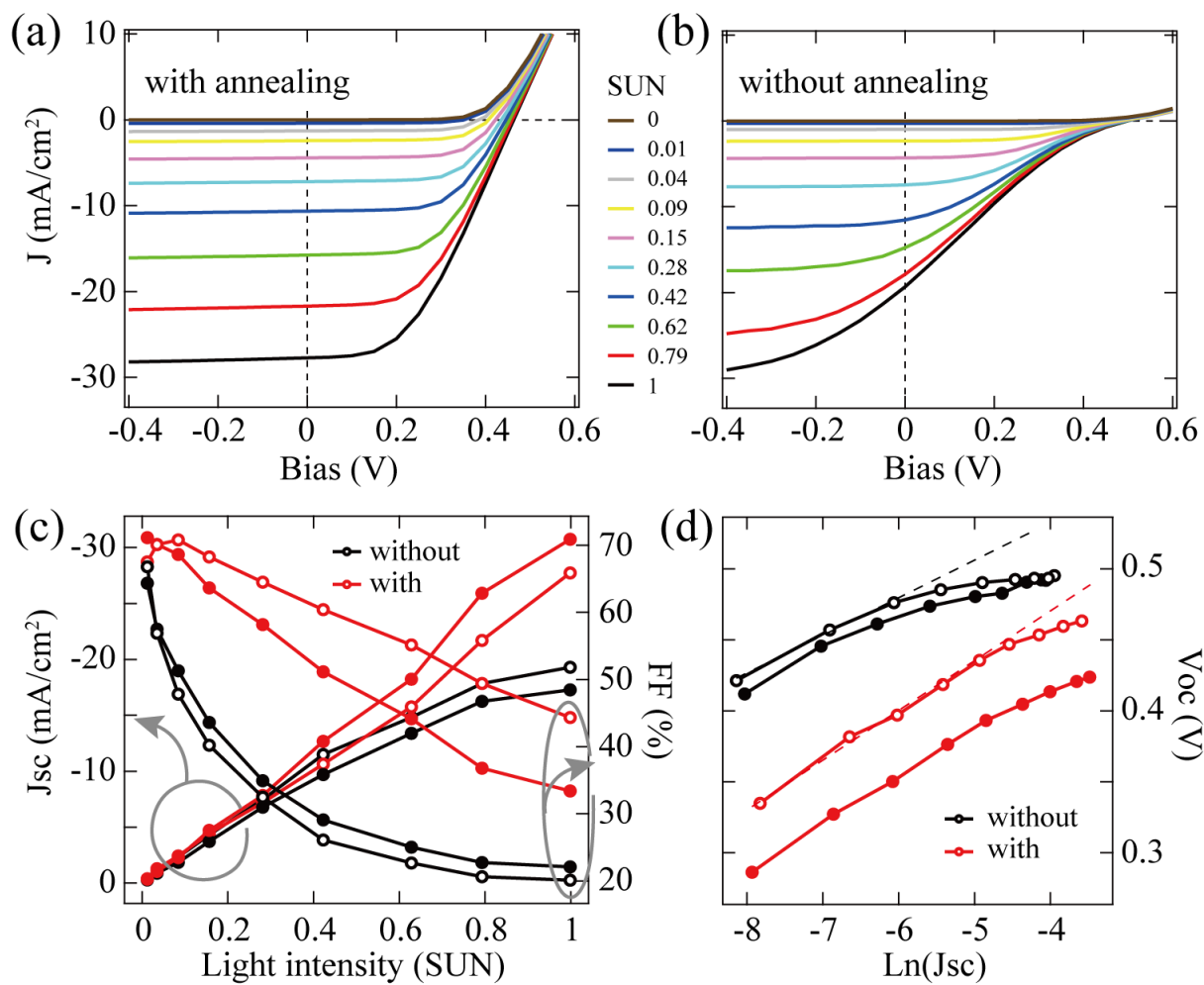

**Fig. S2.** Light intensity dependance of the *J-V* characteristics of (a) the w-a and (b) wo-a GSSCs 1 h after the hole doping. (c) Plot of the $J_{SC}$ and *FF* as a function of the light intensity. (d) Plot of $\ln J_{SC}$ vs. $V_{OC}$. For the sake of comparison, the plots of the pristine GSSCs were added in (c) and (d) (solid circles). The dashed lines are guide to the eye.

**Table. S1.** Photovoltaic parameters of the pristine GSSCs.

| Pristine | $J_{SC}$ (mA/cm$^2$) | $V_{OC}$ (V) | FF (%) | PCE (%) | n | $R_s$ (Ω) | SBH (eV) |
|---|---|---|---|---|---|---|---|
| w-a | 29.2 | 0.41 | 33 | 3.9 | 1.3 | 4.7 | 0.82 |
| wo-a | 14.7 | 0.47 | 22 | 1.5 | 1.6 | 7.9 | 0.89 |

**Table. S2.** Photovoltaic parameters of the pristine GSSCs 7 days after fabrication.

| Pristine | $J_{SC}$ (mA/cm$^2$) | $V_{OC}$ (V) | FF (%) | PCE (%) | n | $R_s$ (Ω) | SBH (eV) |
|---|---|---|---|---|---|---|---|
| w-a | 29.4 | 0.42 | 33 | 4.0 | 1.3 | 4.7 | 0.86 |
| wo-a | 12.6 | 0.45 | 20 | 1.2 | 1.5 | 10.1 | 0.92 |

**Table. S3.** Photovoltaic parameters of the GSSCs shortly after hole doping.

| HNO$_3$ doping | $J_{SC}$ (mA/cm$^2$) | $V_{OC}$ (V) | FF (%) | PCE (%) | n | $R_s$ (Ω) | SBH (eV) |
|---|---|---|---|---|---|---|---|
| w-a | 27.8 | 0.48 | 53 | 7.1 | 1.5 | 3.5 | 0.86 |
| wo-a | 25.5 | 0.51 | 19 | 2.5 | 1.5 | 6.5 | 0.93 |

**Table. S4.** Photovoltaic parameters of the GSSCs 1.5 hours after hole doping. The *n*, *R*s, *SBH* were not extracted due to the absence of the dark *J-V* curve.

| HNO$_3$ doping | $J_{SC}$ (mA/cm$^2$) | $V_{OC}$ (V) | FF (%) | PCE (%) | n | $R_s$ (Ω) | SBH (eV) |
|---|---|---|---|---|---|---|---|
| w-a | 30.7 | 0.45 | 40 | 5.6 | - | - | - |
| wo-a | 17.0 | 0.47 | 19 | 1.5 | - | - | - |

**Table. S5.** Photovoltaic parameters of the GSSCs 7 days after hole doping.

| HNO$_3$ doping | $J_{SC}$ (mA/cm$^2$) | $V_{OC}$ (V) | FF (%) | PCE (%) | n | $R_s$ (Ω) | SBH (eV) |
|---|---|---|---|---|---|---|---|
| w-a | 27.1 | 0.45 | 37 | 4.5 | 1.5 | 4.1 | 0.85 |
| wo-a | 14.5 | 0.48 | 18 | 1.2 | 1.6 | 8.7 | 0.91 |